\documentclass[11pt, aps,prd,twocolumn, showpacs,showkeys]{revtex4-1}
\usepackage{graphicx}
\usepackage{amsmath}
\usepackage{amsfonts}
\usepackage{amssymb}
\begin{document}
\preprint{}
\title{Weakly Charged Compact Stars in $f \left( R \right)$ gravity}
\author{H. Mansour}
\email[mansourhouda48@gmail.com]
\thanks{}
\altaffiliation{}
\affiliation{Laboratoire de Physique Fondamentale et Appliqu\'{e}e, Blida University, P.O.Box 270, Route de Soumaa, Blida, 09000 Algeria}
\author{B. Si Lakhal}
\email[contact: bahia.silakhal@g.enp.edu.dz]
\thanks{}
\altaffiliation{}
\affiliation{Preparatory Classes Department, National Polytechnic School, 10 Hassen Badi Street, El Harrach, Algiers, 16200 Algeria}
\author{A. Yanallah}
\email[yanallahabdelkader@hotmail.com]
\thanks{}
\altaffiliation{}
\affiliation{Physics Department, Faculty of Sciences, Blida University, P.O.Box 270, Route de Soumaa, Blida, 09000 Algeria}
\date{\today}

\begin{abstract}
We study electrically charged compact stars in the framework of extended theory of gravity (ETG). We assume that the charge density is proportional to the energy density. The polytropic equation of state is chosen to describe the state of the charged perfect fluid. We aim to find the Oppenheimer Volkoff (OV) mass limit for charged compact stars. A detailed numerical study is performed. We show the dependence of the mass-radius diagram of the spheres on the values of the perturbatif parameter $\beta$, the polytropic exponent $\gamma$ and the charge fraction $\alpha$. Our results are compared with those found in the literature in the case of applying General Relativity (GR).
\end{abstract}
\maketitle
\section{Introduction}
Just after the discovery of the accelerating expansion of the universe in 1998, extended theory of gravity (ETG) has shown an intensive interest in the literature \cite{Cappoziello1, Cappoziello2, nojiri, Palatini}. It is an approach that seeks to cure the shortcomings of GR but in the same time owes to preserve their positive results, instead of introducing unknown fluid's components in the universe that seem artificial (dark matter, dark energy\dots etc). Another motivation of ETG is to solve the problems coming out at ultra-violet and infra-red scales. The idea of ETG consists in extending the Einstein's theory of GR by considering a nonlinear function of the Ricci scalar curvature in the action. For historical review, see \cite{Schmidt}. In fact the idea of ETG started earlier just after the advent of GR's theory of gravity \cite{Weyl1919,Eddigton1923}, because of its non-renormalisability that makes it irrelevant as a quantum theory. So in 1962 Utiyama and DeWitt \cite{Utiyama-DeWitt} showed that renormalization at one-loop requires that the Einstein-Hilbert action be supplemented by higher order curvature terms \cite{Utiyama}. The form of the function $f(R)$ is not specified, but is constrained to keep the positive GR results at Solar System scales. Furthermore, to preserve the other correct results of GR, criteria for viability of ETG must be applied namely: correct cosmological dynamics, stability, absence of ghosts, correct Newtonian and post-Newtonian limit, well-settled Cauchy problem and cosmological perturbations compatibility with the cosmic microwave background radiation and large-scale structures \cite{Cappoziello1}.

In general, Alternative Theories of Gravity are tested by studying the formation and the evolution of stars considered as suitable test-beds. In that context, we aim to study the properties of electrically charged compact stars in the framework of $f(R)$ gravity. In fact, compact stars are well studied objects in the literature: Chandrasekhar \cite{Chandrasekhar1939} showed that white dwarfs are compact stars in which quantum degeneracy of the electrons is responsible for their stability. These stars are cold and as their configuration gets more compact, the electrons get more and more relativistic and the radius will tend almost to zero (few kilometers for some neutron stars). It is worth mentionning, here, that there exists a limit to the radius of the sphere which is related to the mass of the star called the Chandrasekhar limit, generally taken as  $1.44\,M_{\odot}$. Landau in Ref. \cite{Landau}, through heuristic arguments, found that the mass limit for a white dwarf is $\sim 1\, M_{\odot}$. He also deduced that the radius of the star could be $\sim5000\,km$ . 
Tolman \cite{Tolman1939} and Oppenheimer and Volkoff \cite{Oppenheimer1939} showed that neutron stars, much more compact than white dwarfs, have also equilibrium configurations and that these neutron stars have a mass limit, called the Oppenheimer-Volkoff limit. However, this limit depend on the equation of state used. As stated in Ref. \cite{Cappoziello4}, this limit lie in the range $1.4\, M_{\odot}$ to $6\, M_{\odot}$. Note that the Chandrasekhar limit which appears within Newtonian gravitation, turns into the Oppenheimer-Volkoff limit in the degenerate matter, when relativistic kinematic effects become significant.

The stability of charged fluid spheres was first studied by Bekenstein \cite{Bekenstein}, then followed by many authors \cite{Zhang, Felice, Anninos, Ray, Ghezzi, Siffert, Lemos2008, Lemos2010, Lemos2013}. 

This paper is outlined as follows: The basic equations in GR needed for the purpose of our work, namely the Einstein-Maxwell equations for a charged sphere, in static spherical symmetry case are given in section II. We then briefly recall some properties of the $f(R)$ gravity in section III. Then in section IV, to get a closed system of equations, we define a polytropic equation of state and a charge density profile. Then we explicitely write down the set of equations of the charged compact stars. Section V is devoted to the numerical study of compact charged stars. With the chosen form for the $f(R)$ function, we study the dependence of radius to mass ratio on the perturbatif parameter $\beta$ for some fixed values of polytropic exponent $\gamma$ and small charge fraction $\alpha$. Then we deduce The OV limit. In section VI we discuss our results. Finaly we conclude in section VII. This paper is endowed with an appendix dedicated to dimensionless form of the set of equations needed for calculation program, for the purpose of simplification.
\section{Basic equations in general relativity}
We suppose that the metric is spherically symmetric. The line element is then assumed to be of the form:
\begin{equation}
\nonumber
{{\it ds}}^{2}=-B\left(r\right){{\it dt}}^{2}+A\left(r\right){{\it dr}}^{2}+{r}^{2}{d\theta}^{2}+{r}^{2}\sin^{2}\!\theta{d\varphi}^{2},
\end{equation}
where $(t,r, \theta, \varphi)$ are the like-Schwarzchild coordinates. $A(r)$ and $B(r)$ depend only on $r$ to ensure that we deal only with static configurations.
For the seek of simplicity, we put
\begin{equation}
a(r)=A^{-1}(r)=1-\dfrac{2 m(r)}{r}+\dfrac{q(r)^{2}}{r^{2}},      
\end{equation} 
in the metric, where we have introduced the mass $m(r)$ and the charge $q(r)$ inside a star's shell of radius $r$. 
The Einstein-Maxwell equations in the presence of electrically charged matter are:
\begin{equation}
G_{{\mu \nu }}=- \frac{8\pi G}{c^{4}}\ T_{{\mu \nu }}, 
\end{equation}
\begin{equation}
\nabla _{\nu }F^{\nu \mu }=4\pi j^{\mu},
\end{equation}
where the Greek indices $\mu$ and $\nu$ run from $0$ to $3$. The Einstein tensor is defined as
\begin{equation}
G_{{\mu \nu }}=R_{{\mu \nu }}-\frac{1}{2}\,Rg_{{\mu \nu }},
\end{equation}
where $g_{{\mu \nu }}$ is the metric tensor, $R_{{\mu \nu }}$  is the Ricci tensor and $R$ is the Ricci scalar. We assume that the interior of the star is filled with a perfect fluid and radiation so that, the energy-momentum tensor $T_{{\mu \nu }}$ is given by:
\begin{equation}
T_{{\mu \nu }}=E_{{\mu \nu }}+M_{{\mu \nu }}.
\end{equation}
$M_{{\mu \nu }}$ stands for the energy-momentum tensor of a perfect fluid
\begin{equation}
M_{{\mu \nu }}=pg_{{\mu \nu }}+ \left( p+\rho \right) u_{{\mu}}u_{{\nu}},
\end{equation}
where $\rho$ is the energy density and $p$ the pressure. $u_{{\mu}}$ is the four-vector velocity of the fluid (with $u_{\mu}u^{\mu}=-1$). The electric current density $j^{\mu}$ is related to the electric charge density $\rho_{e}$ by the following equation
\begin{equation}
j^{\mu}= \rho_{e} u^{\mu}.
\end{equation}
$E_{{\mu \nu }}$ is the electromagnetic energy-momentum tensor given by:
\begin{equation}
E_{\mu \nu }=\frac{1}{4\pi }\left( F_{\mu }^{\gamma }F_{\nu \gamma }-\frac{1%
}{4}g_{\mu \nu }F_{\gamma \beta }F^{\gamma \beta }\right),
\end{equation}
where
\begin{equation}
F_{\mu \nu }=\partial _{\mu }A_{\upsilon }-\partial _{\nu }A_{\mu }.
\end{equation}
is the electromagnetic Faraday-Maxwell tensor.
As we assume a static spherically symmetric electric field, the only nonvanishing components of the $F_{\mu\nu}$ tensor are $F_{01}=-F_{10}$. 
In what follow, we will adopt the geometrical units $c=G=1$, and for the seek of simplicity, we will drop the radial dependence from all functions of $r$.
\section{Extended theory of gravity}
In this study we intend to examine the quadratic corrections,
\begin{equation}
f(R)=R+\frac{1}{2}\beta R^{2},
\end{equation}
to the Hilbert-Einstein action, because it has shown consistent results for some cosmological phenomena. $\beta$ is a parameter of the quadratic corrections to the Ricci scalar. According to \cite{starobinsky}, the stability conditions are: $f'(R)>0$ and $f"(R)>0$ which leads to $\beta>0$. Hence, the conditions of stability guaranty the attractive nature of gravitational interaction and the absence of tachyons. We give here a brief review of the formalism of ETG in the framework of variational principle of least action. ETG is based on the variation of the metric and matter action $A$:
\begin{eqnarray}
\delta A=\delta \left[ \int d^{4}x\sqrt{-g}\left( f(R)+\chi {\cal L}_{m}\right)\right] =\delta S+\delta S_{m}.\
\nonumber
\end{eqnarray}
${\cal L}_{m}$ is the minimally coupled ordinary matter Lagrangian density. The variation of the modified Einstein-Hilbert action is given by
\begin{eqnarray}
&&\delta S \!\! =\!\! \delta\!\int\! d^{4}x\sqrt{-g}f(R)\!\!=\!\!\int\!\! d^{4}x\!\left\lbrace \sqrt{-g}[F(R) R_{\mu \nu }\right.\nonumber \\ \nonumber  &&\qquad -\frac{1}{2} f(R) g_{\mu\nu}] +[g_{\mu\nu}\partial^{\sigma}\partial_{\sigma}(\sqrt{-g}F(R))\\
&&\qquad - \left. g_{\sigma\nu}\partial^{\sigma}\partial_{\mu}(\sqrt{-g}F(R))]\right\rbrace \delta g^{\mu\nu},
\end{eqnarray}
with $F\left( R\right)=\frac{df(R)}{dR}$ and $\chi=8\pi$ in geometrical units. $g$ is the determinant of the metric tensor $g_{\mu\nu}$.
Then the variation of the action with respect to $g_{\mu \nu }$ gives the following gravitational field equations
\begin{eqnarray}
F\left( R\right) R_{\mu \nu }&-&\nabla _{\nu }\nabla _{\mu }F\left( R\right)+g_{\mu \nu }\square F\left( R\right)\quad \label{117} \\ &&\qquad\qquad -\frac{1}{2}g_{\mu \nu }f(R)=-\chi T_{\mu \nu },\nonumber
\end{eqnarray}
with $\nabla _{\nu }$ is the covariant derivative, $\square$ is the D'Alembert operator. 
The trace of equation (\ref{117}) is
\begin{equation}
R F(R)+ 3\square F(R)-2 f(R)=-\chi T.
\end{equation}
With our chosen function $f(R)$, and following Ref. \cite{Cappoziello2}, the field equations are found to be
\begin{eqnarray} \label{equ1}
 && G_{\mu\nu}+{\beta}\left(R_{\mu\nu}-g_{\mu\nu}\,R/4\right.\\
 &&\qquad \qquad +\left. g_{\mu\nu}\square -\triangledown_{\mu} \triangledown_{\nu} \right) R =-\chi T_{{\mu \nu }}. 
 \nonumber
\end{eqnarray}  
\section{The equations of equilibrium in ETG}
The expression (\ref{equ1}) gives the equations which extend those found in GR as in Ref. \cite{arbanil}. These equations, together with the Maxwell equation, can be written explicitely as: 
\begin{equation}
{\frac {\rm dq}{{\rm d}r}} =\frac{4\,\pi\,\rho_{{e}}
 {r}^{2}}{\sqrt {a}},
 \label{maxwell}
\end{equation}   
\begin{eqnarray}
\nonumber
&& r^{2}\left(  \frac{2 a R'}{r}+a R''+\frac{a' R'}{2} -\frac{R^{2}}{4} +\frac{a  R B''}{2 B}  \right.\\
\nonumber && \qquad \left. +\frac{a'R B'}{4 B}-\frac{a R B'^{2}}{4 B^{2}}+ {\frac{a R B'}{r B} }\right)\beta +\\  && \qquad\qquad 1-a-a' r =8 \pi r^{2} \rho_{_{m}}+\frac{q^{2}}{r^{2}},\label{Eins1} 
\end{eqnarray}
\begin{eqnarray}
\nonumber
&& r^{2}\left(\frac{R^{2}}{4} -\frac{a R B''}{2 B} -\frac{a' R B'}{4 B}+\frac{a R B'^{2}}{4 B^{2}} \right.\\
\nonumber &&  \qquad\left. -\frac{a R' B}{2 B} - \frac{a'R}{r}- \frac{2 a R'}{r}\right) \beta- \\\label{Eins2}
&& \quad\quad 1+a + \frac{a r B'}{B}=8 \pi r^{2} p-\frac{q^{2}}{r^{2}},
\end{eqnarray}
\begin{eqnarray}
\nonumber
&& r^{2}\left(\frac{R^{2}}{4} -a R''-\frac{a'R'}{2} -\frac{a R' B'}{2 B}+\frac{R}{r^{2}} -\frac{a'R}{2r} \right.\\
\nonumber && \left. -\frac{aRB'}{2rB}-\frac{a R'}{r} -\frac{a R}{r^{2}} \right) \beta+r^{2}\left(\frac{a B''}{2 B} +\frac{a' B'}{4 B} \right.\\
 && \left.-\frac{a B'^{2}}{4 B^{2}}\right)+r\left(\frac{a'}{2}-\frac{a B'}{2 B} \right) =8 \pi r^{2} p+\frac{q^{2}}{r^{2}} \label{Eins3}
\end{eqnarray}
Note that the prime symbol $'$ stands for a derivative with respect to $r$.
\subsection{Equations of state}
In this work we assume a polytropic relation between the pressure $p$ and the energy density $\rho$. Therefore we choose 
\begin{equation}
p \left( r \right) =\omega\, \left( \rho \left( r \right) 
 \right) ^{\gamma}\label{polytr}
\end{equation}
as an equation of state, where $\omega$ is the polytropic factor and $\gamma$ is the polytropic exponent. 
This is in order to compare our results found in the context of ETG with those found in GR, particulary with those given in \cite{arbanil}.

Note that, often in the literature, the polytropic index $n$ is used, instead of the polytropic exponent $\gamma$ related by: $\gamma=1+1/n$.  
We also assume, for the seek of simplicity, that the charge density is proportional to the energy density:
\begin{equation}
\rho_{e}\left( r\right) =\alpha\rho\left( r\right),\label{equ-stat2} 
\end{equation}
where $\alpha$ is the charge fraction which is dimensionless in geometrical units. 
By using Eqs. (\ref{polytr}-\ref{equ-stat2}) together with the three Eqs. (\ref{Eins1}-\ref{Eins3}) in ETG and the Maxwell equation, we have a system of six equations to solve in order to find the six unknown variables: $q(r),\; m(r),\; B(r),\; \rho(r),\; \rho_{e}(r)$ and $p(r)$.
Moreover, to get suitable system to solve numerically, we adopt the following equations system, deduced from Eqs. (\ref{maxwell}-\ref{Eins3}) in such a way to eliminate the second derivative of $R$:
\begin{eqnarray}
&& \frac{dq}{dr}=\frac{4 \pi \alpha r^{2}}{\sqrt{a}}\left(\frac{p}{\omega} \right)^{\frac{1}{\gamma}},\\
&& m(r)=\frac{1}{2}\left(1-a r \right)r+\frac{1}{2}\frac{{q^{2}}}{r}, \\
&& \frac{dp}{dr}=-\frac{1}{2}wp-\left(\frac{p}{\omega} \right)^{\frac{1}{\gamma}}\left( \frac{w}{2} -\frac{\alpha q}{r^{2}\sqrt{a}}\right),\\
&&\nonumber\\
\label{192}
\nonumber && \frac{dR}{dr}=\frac{1}{2 r^{3}\beta a\left(wr+4 \right)}\left[4r^{2}\left(R\beta+1 \right)\left( wr+1\right)a
\right.\\
 && \quad\left.
-\beta R^{2} r^{4}-32\pi p r^{4}-4\beta R r^{2}-4r^{2}+4 q^{2}\right], \\
&&\nonumber\\
\nonumber
&&\frac{dw}{dr}=\frac{1}{12 ar^{4}\left(\beta R+1 \right) }\left[-3\beta r^{5}w R^{2}-2 r^{5} Rw\right.\\
 && \left.+32\pi r^{5} w \left(\frac{p}{\omega}
\right)^{\frac{1}{\gamma}} -12 \beta r^{3} R w(a+1)-12 a r^{3} w\right. \nonumber\\
 && \qquad \left. +12 r w q^{2}
-12 w r^{3}+6\beta r^{4} R^{2}+192\pi p r^{4}\right. \nonumber\\
 && \qquad\left. +128\pi r^{4}\left(\frac{p}{\omega} \right)^{\frac{1}{\gamma}}+4 r^{4}R+24 q^{2}\right],
\end{eqnarray}
\begin{eqnarray}
\nonumber
&& \frac{da}{dr}=\!-\!\left[32 \pi r^{4}\left( wr\!+\!4\right)\left(\frac{p}{\omega} \right)^{\frac{1}{\gamma}}\!+\!6 r^{4} a \left(\beta R\!+\!1 \right)w^{2}\right. \nonumber \\
&& \quad \left. -3r w \left(\beta r^{4} R^{2}+\frac{2}{3}r^{2} R (r^{2}-6\beta a+6\beta )\right.\right. \nonumber 
\\
 &&\quad \left. \left. -4 (a r^{2}- r^{2}+ q^{2})\right) -6\beta r^{4} R^{2}\right. \nonumber \\
&&\quad \left.-8 r^{2} R\left(r^{2}-3\beta a+3\beta \right) +192\pi p r^{4}+24( a r^{2}\right. \nonumber \\
&&\quad  \left.- r^{2}+ q^{2})\right]\!\!\times\!\!\frac{1}{6 r^{3}\left( wr+4\right)\left(\beta R+1 \right)},
\end{eqnarray}
where 
\begin{equation}
\nonumber
w=\frac{d\ln B}{dr}.
\end{equation}
Here, we have to mention that $\beta$ should be different from $0$ in Eq. (\ref{192}) as a consequence of recombinations of our differential equations system. However, if we want to recover the results found in GR case \cite{arbanil}, we have to recompile the program from the initial set of Eqs. (\ref{Eins1}-\ref{Eins3}).
The stellar structure differential equations are integrated numerically from the center to the surface of the star. We need to set up initial and boundary conditions, so we put: $m(r=0)=0$, $q(r=0)=0$, $p(r=0)=p_{cr}$, $\rho(r=0)=\rho_{cr}$, $\rho_{e}(r=0)=\rho_{ecr}$. Before proceeding to numerical computations, our set of differential equations are transformed to non-dimensionless form (see Eqs. (\ref{262}-\ref{310}) in appendix) and supplemented with the following relevant boundary conditions: $ u(0) = 31.62,\; \theta(0) = 1,\; a(0) = 1,\; w(0) = 1,\; R(0) = 0$. Note that the initial value of $u(\epsilon)$ can take any arbitrary value due to the form of the equation: $q(r)=\frac{\epsilon^2 u(\epsilon)}{\sqrt{4\pi \rho_{cr}}}$. Moreover, due to the stiffnes of equations and to avoid singularity problems when integrating (due particularly to the limits of the machine), the initial conditions are started slightly above the center of the sphere.\\
The radius $R_{s}$ of the compact star is reached when we get from the numerical program $p(R_s)=0$ at the surface of the star. The interior solution is smoothly connected to the exterior Reissner-Nordstr\"{o}m metric through the equation of boundary condition
\begin{eqnarray}
 a( R_{s}) & =&\frac{1}{A( R_{s})  }\nonumber\\
&=& 1-\frac{2 M}{R_{s}}+\frac{Q^{2}}{R_{s}^{2}}=-B( R_{s}),
\end{eqnarray} 
with $M$ being the total mass of the compact star and $Q$ its total charge. 
\section{Simulation and results}
Following Ref. \cite{arbanil}, we restore the value of the gravitational constant to $G=7.42611\times 10^{-28}m/kg$, keeping $c=1$, in order to get results in suitable units. To solve the set of differential equations, we use The 4th order Range-Kutta method and for some stiff cases  we use the Rosenbrock method provided by Maple 2016 software.

In our numerical program, we choose the central energies in the range $[10^{10}, 10^{19}] kg/m^{3}$ for more easier comparison with other works in the literature related to some types of compact stars, namely white dwarfs, neutron stars and black holes. Furthermore, the charge fractions that we consider throughout this work correspond to small charges. The restriction of the charge fraction $\alpha$, in our work, to small values is justified as follows: 1) We think that if strongly charged stars exist in the universe, then their electric felds will be observed on the surrounding of the stars. 2) The second reason is purely technical due to the occuring of singularity problems when executing our numerical program for high values of $\alpha$. As pointed out in different works related to charged compact stars \citep{Bekenstein,arbanil, Ray1}, these objects are not stable and collapse to charged black holes; but we expect that small charges would not affect the stability of such objects. \\
Following again Ref. \citep{arbanil}, we normalized the polytropic constant $\omega$ in such a way that it turns out a function of the polytropic exponent $\gamma$: $\omega(\gamma)=1.47518\times10^{-3}\,({ 1.78266\times 10^{15} })^{1-\gamma}$. The normalised energy density is then $\rho_{0}= 1.78266\times 10^{15}kg/m^{3}$ corresponding to a pressure $p_{0}=2.62974\times 10^{15}kg/m^{3}$.

\begin{figure*}
a)\includegraphics[scale=0.35]{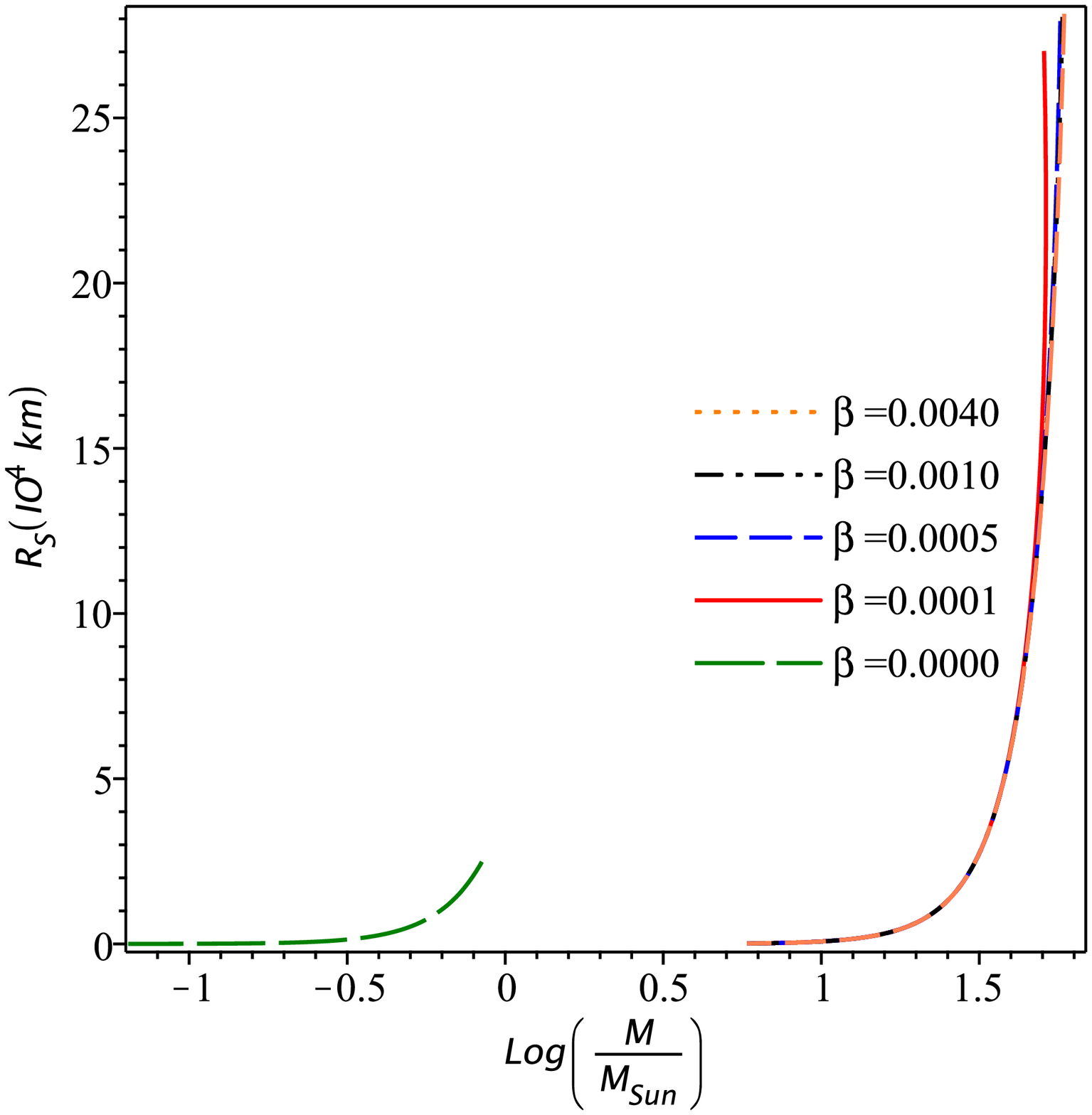}
b)\includegraphics[scale=0.35]{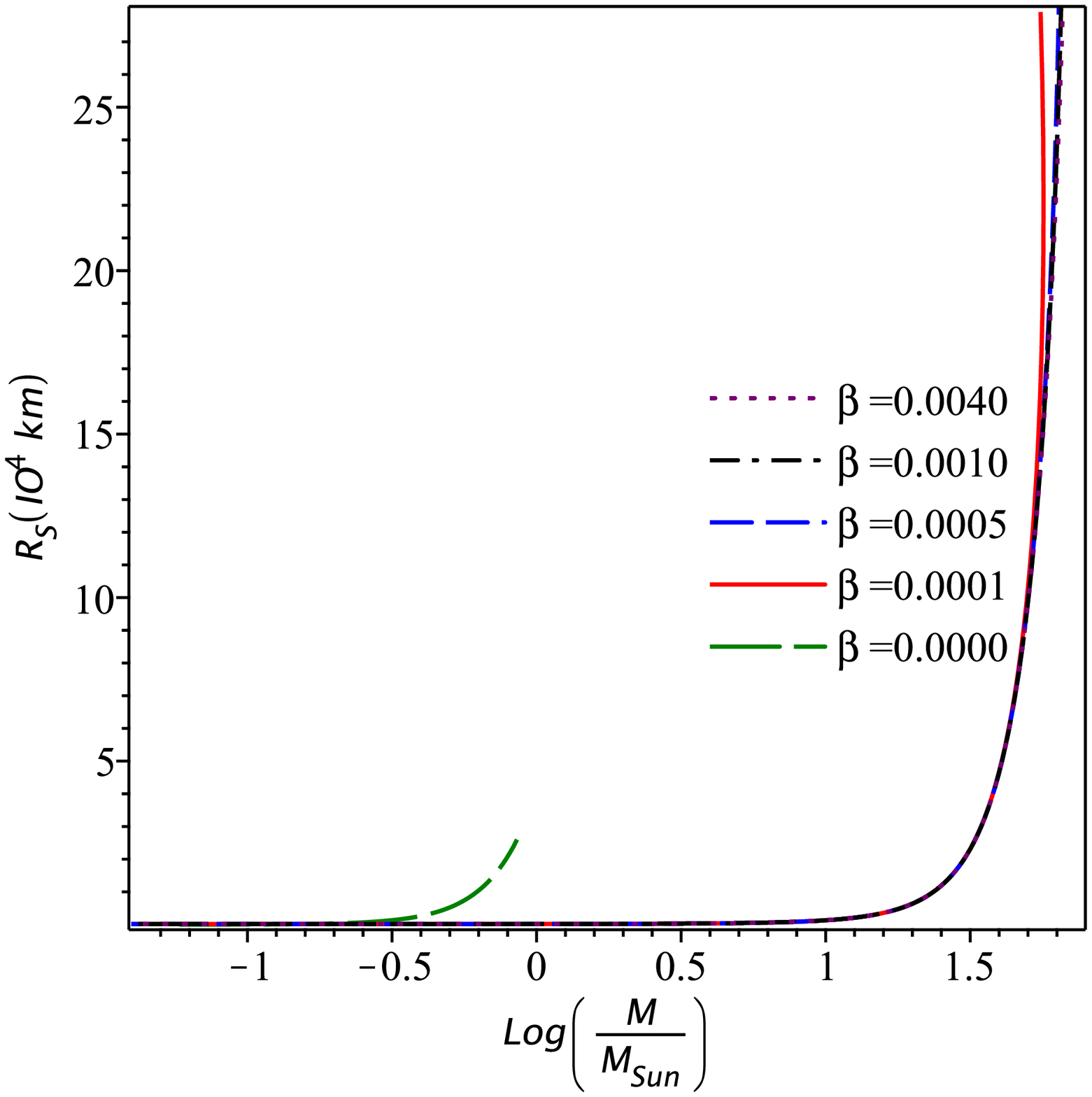}
a1)\includegraphics[scale=0.35]{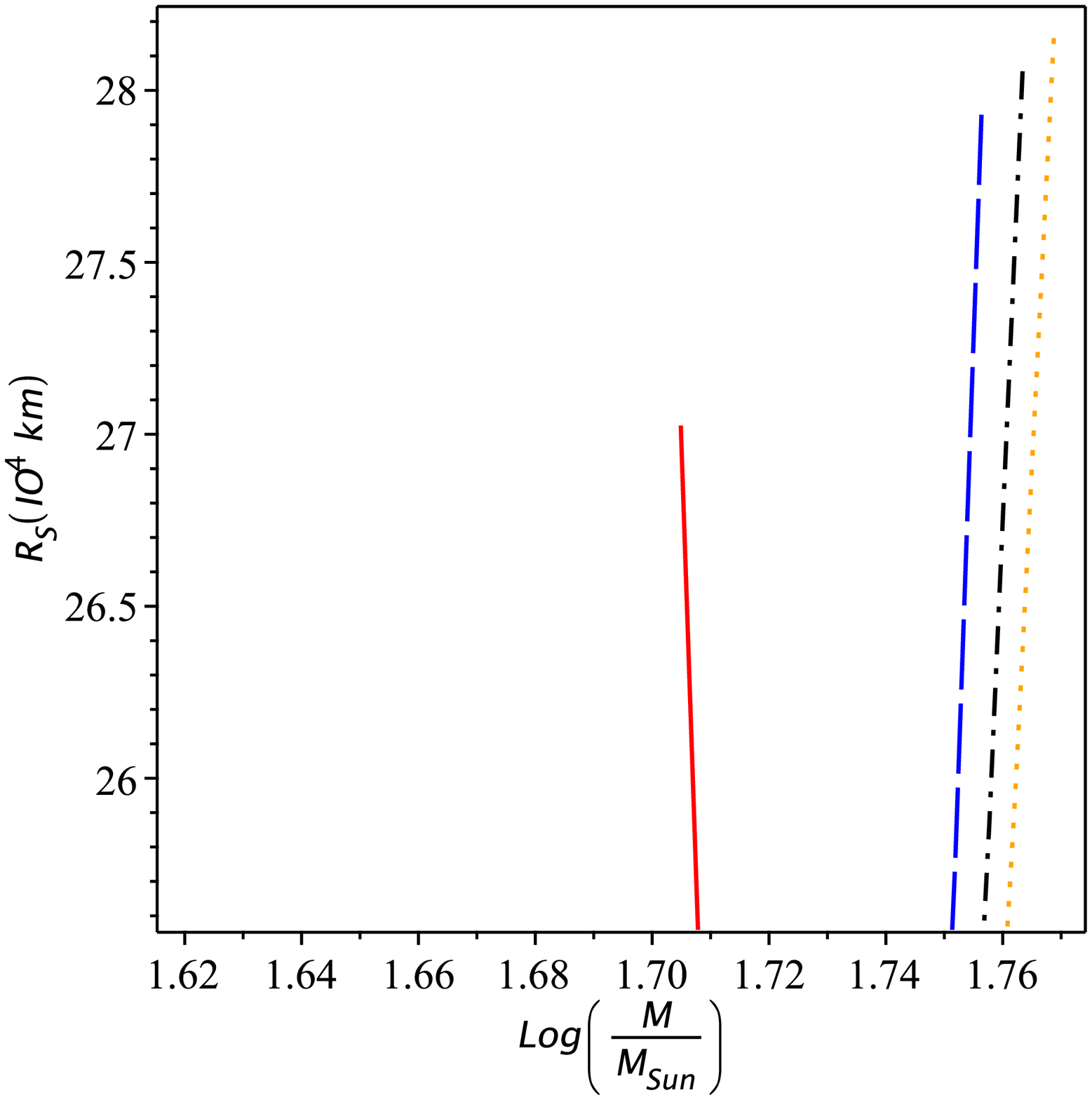} 
b1)\includegraphics[scale=0.35]{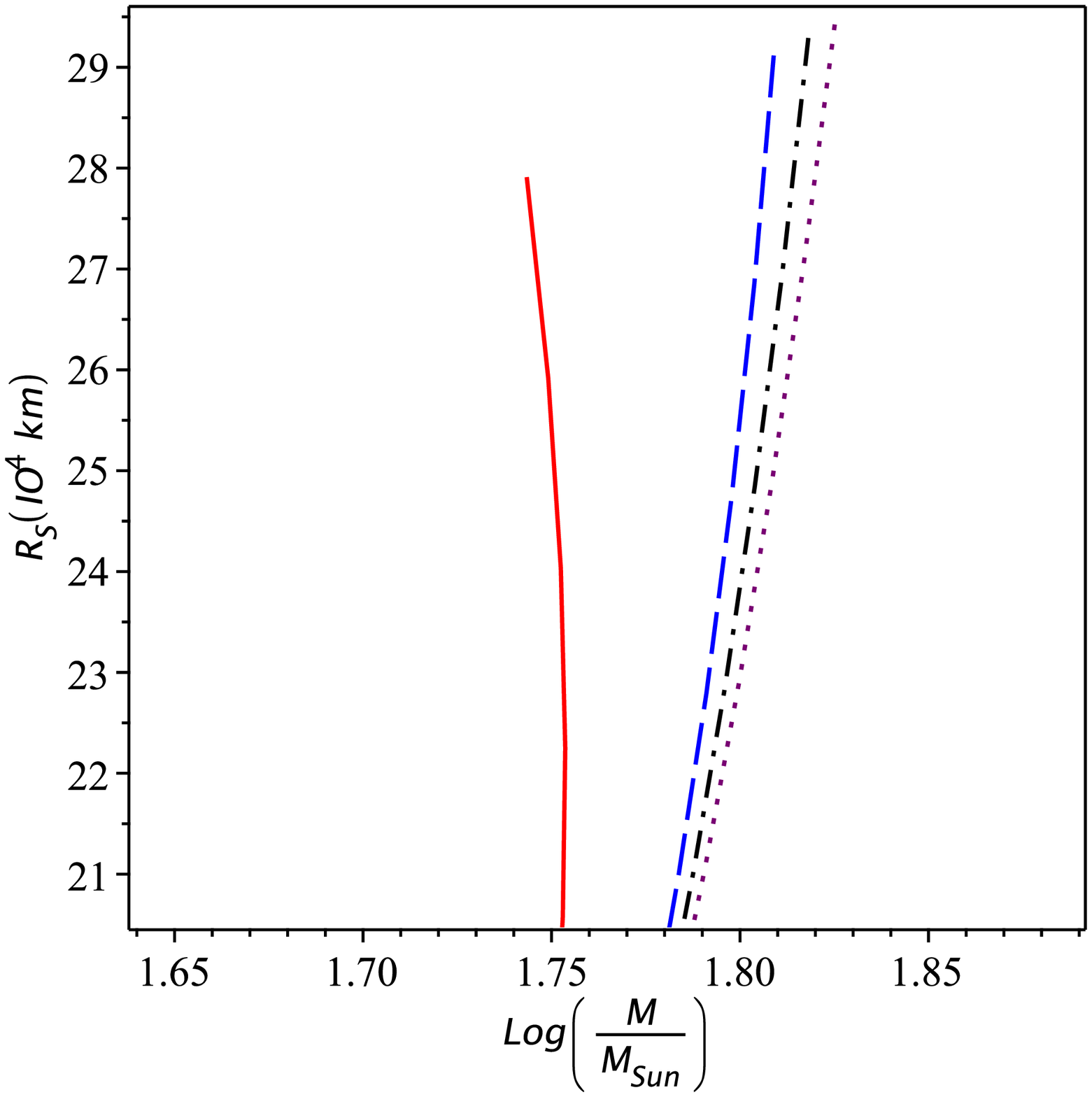} 
\caption{\label{fig:epsart1}(color online). The mass-radius diagram of the charged polytropic sphere for different values of the perturbatif parameter $\beta$ (including the GR case). We considered here the case of $\gamma=5/4$ and respectively for a)$\alpha=0.001$ and for b)$\alpha=0.006$, taking the central energy density $\rho_{cr}$ in the range $\left[ 10^{10}, 10^{19}\right]kg/m^{3}$. a1) and b1) represent respectively the magnification of a) and b) on their right top corners. Clearly these figures show a behavior transition from GR sector to ETG sector.}
\end{figure*}

Figs.\ref{fig:epsart1}.a and \ref{fig:epsart1}.b, give the radius of the charged sphere as a function of its mass $M$ normalised to the Sun's mass $M_{\odot}$ (in logarithmic scale), for fixed polytropic exponent $\gamma=5/4$ and respectively for two values of the charge fraction $\alpha=0.001$ and $\alpha=0.006$. We displayed in the figures only the values of $\beta$ that show a visible effect on the graphs compared to those in GR.
Because of the small effect of the perturbatif parameter $\beta$ on these two figures, we displayed in Figs. \ref{fig:epsart1}.a1 and \ref{fig:epsart1}.b1 the zoomed region of their respective right corners.
It is obvious that the central energy density increases along the curves from right to left, in both figures.

\begin{figure*}
a)\includegraphics[scale=0.35]{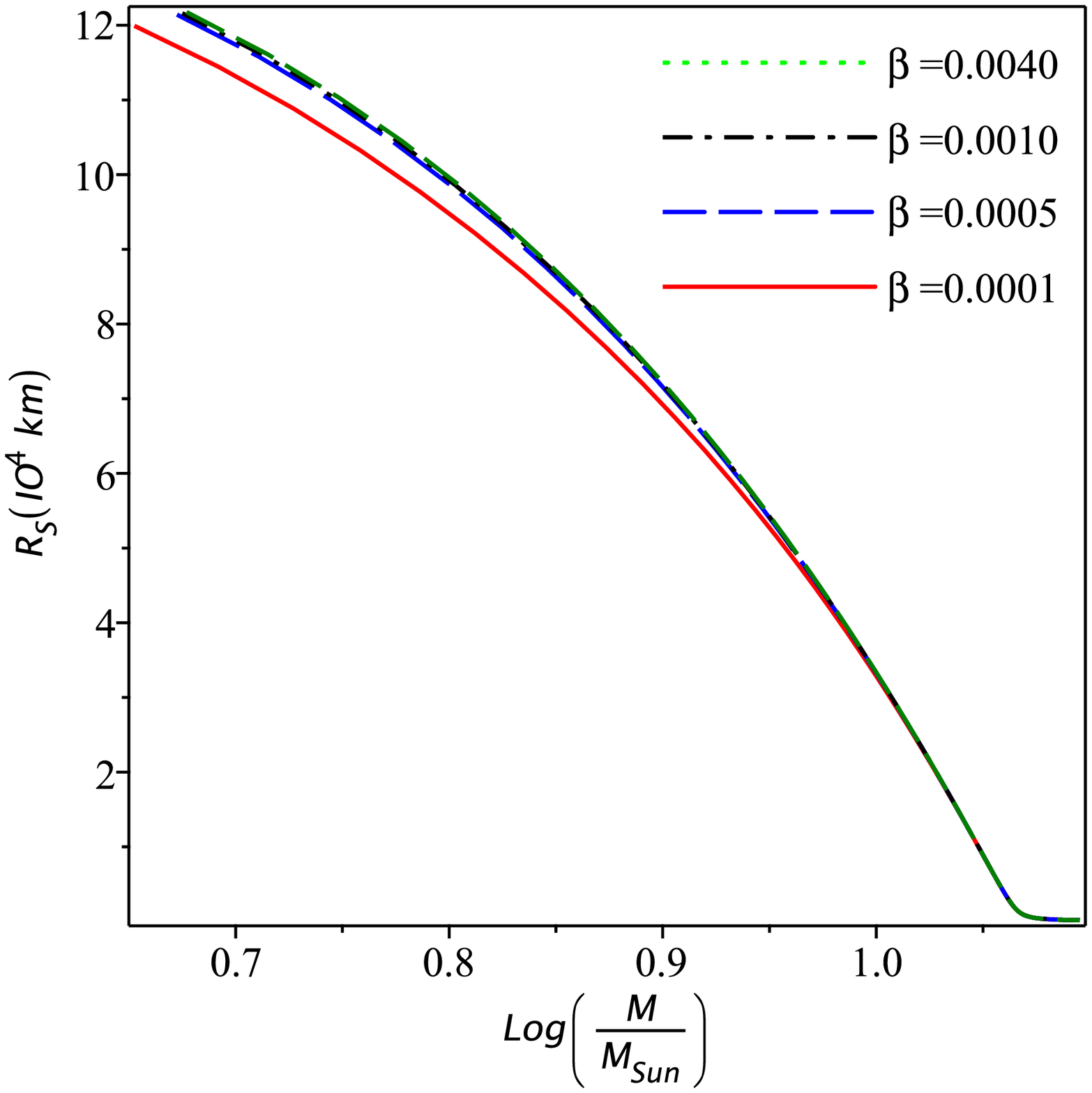}
b)\includegraphics[scale=0.35]{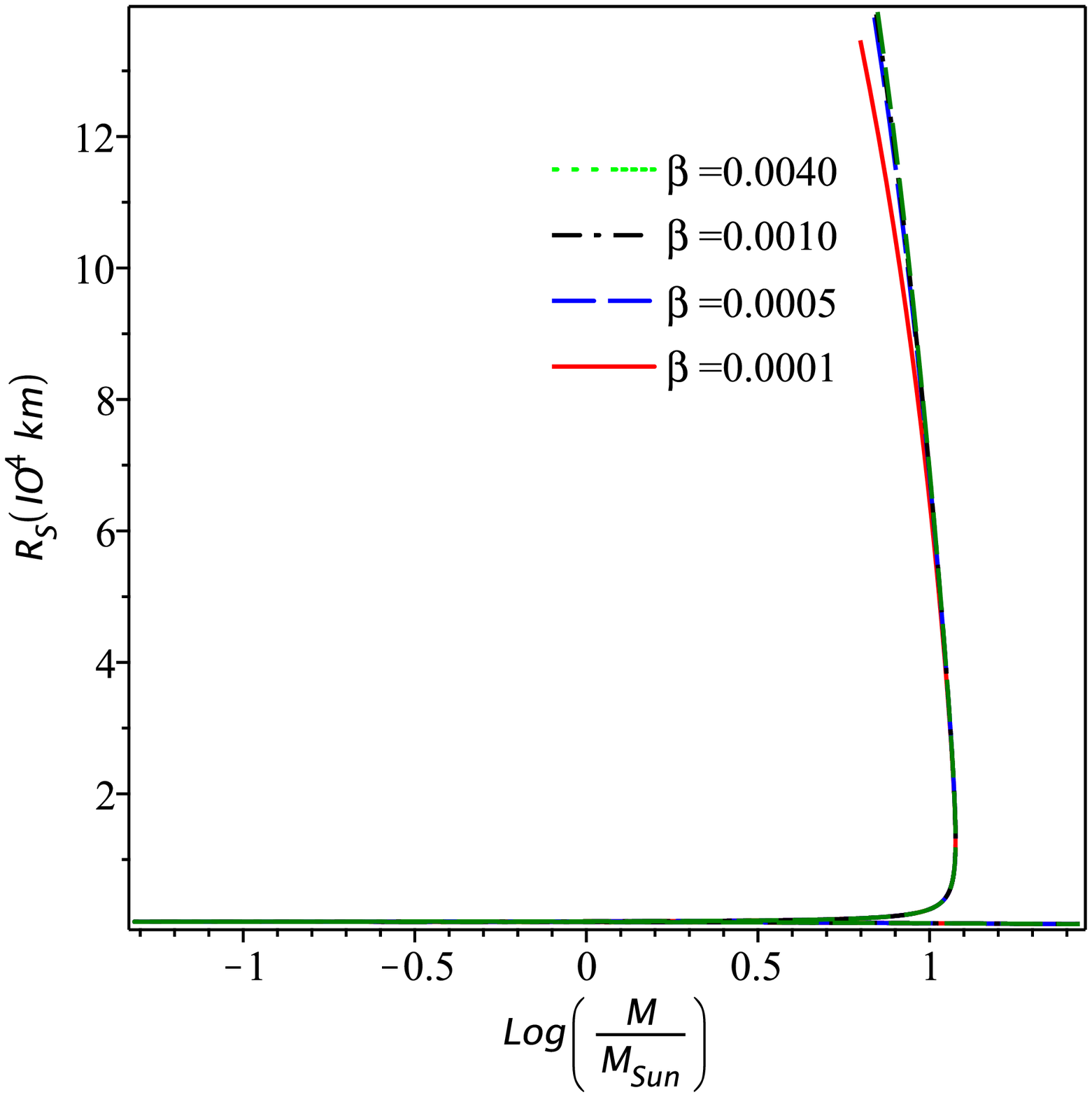}
\caption{\label{fig:epsart2} (color online). The mass-radius diagram of the charged polytropic sphere for different values of  the perturbatif parameter $\beta$ for $\gamma=4/3$, for a) $\alpha=0.001$ and b)$\alpha=0.006$ , taking the central energy density $\rho_{cr}$ in the range $\left[ 10^{10}, 10^{19}\right]kg/m^{3}$.} 
\end{figure*}

Likewise, Figs. \ref{fig:epsart2}.a and ~\ref{fig:epsart2}.b show the radius of the charged sphere as a function of its normalised mass $M/M_{\odot}$ for fixed $\gamma=4/3$ and respectively for two values of the charge fraction $\alpha=0.001$ and $\alpha=0.006$.   
\section{discussion}
Although there exists, in the literature, a number of works related to non charged compact stars and magnetic neutron stars in the framework of $f(R)$ gravity as in Refs. \cite{astashenok1, astashenok2, astashenok3, astashenok4}, to our best knowledge, the study of electrically charged compact stars in $f(R)$ gravity is new and there is a lack of references for comparison with these numerical results, except for the GR case (or $\beta=0$). We notice in Fig. \ref{fig:epsart1}.a, for $\gamma=5/4$, $\alpha=0.001$ and for several displayed values of $\beta$ displayed, that the mass limit of the star is $60$ to $69$ times larger as compared to GR case and the corresponding radius is $9$ to $10$ times heavier (results that we reproduced from \cite{arbanil} by using our numerical program). Whereas, in \ref{fig:epsart1}.b,  when $\alpha$ takes the value $0.006$ but $\gamma$ and $\beta$ keep the same values as in \ref{fig:epsart1}.a, the mass limit is $69$ to $79$ times heavier and the corresponding radius is $10$ to $11$ times larger as compared to those in GR (as seen on the bottom left corner of the two figures).   
We notice that the radius-mass ratio is approximately constant (its value ranges from $0.3$ to $0.5$). This ratio is the OV limit.

Figs.\ref{fig:epsart2}.a and \ref{fig:epsart2}.b show two different behaviors of stars. For the same polytropic exponent $\gamma=4/3$ and for the value of charge fraction $\alpha=0.001$, the radius of the sphere decreases monotically from a value of $12\times10^{4} km$ (corresponding to a mass of $4.7\,M_{\odot}$) to $171km $ (corresponding to a mass of $12.32\,M_{\odot}$) from lower to higher densities. But for $\alpha=0.006$ we see that the radius of the charged sphere drops suddenly to a small value. The OV limit ranges from $0.015$ for Fig. \ref{fig:epsart2}.a to $7.2$ for Fig. \ref{fig:epsart2}.b.

We notice, firstly, that in all the above figures, when $\beta\neq0$, the masses and radii of the spheres show a slightly increasing with the perturbatif parameter $\beta$. Secondly, there is a visible jump of masses and radii of the stars when one skips from GR sector to ETG sector.

Furthermore, in all the $4$ figures (\ref{fig:epsart1}.a, \ref{fig:epsart1}.b, \ref{fig:epsart2}.a and \ref{fig:epsart2}.b), we see that the mass-radius ratio increases with the polytropic exponent, the mass increasing slowly compared to the increasing of the corresponding radius.

\begin{figure}[h]
\includegraphics[scale=0.42]{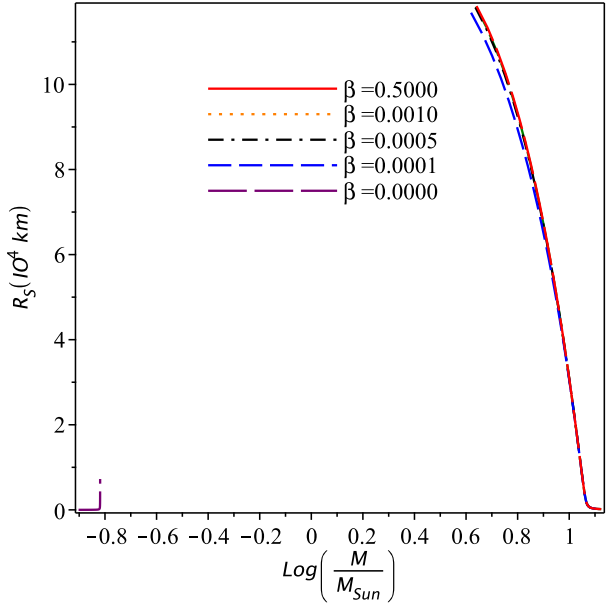}
\caption{\label{fig:epsart3}(color online). The mass-radius diagram of a non charged polytropic sphere for different values of  the perturbation parameter $\beta$ and for $\gamma=4/3$. Note that the GR case is displayed on the bottom left corner of the figure which confirms the transition behavior. The central energy density $\rho_{cr}$ is  in the range $\left[ 10^{10}, 10^{19}\right]kg/m^{3}$.}
\end{figure}

Fig. \ref{fig:epsart3} shows the mass-radius diagram for a non charged polytropic sphere. We notice good agreement with results given in \cite{arbanil} for $\beta=0$, in GR. Non charged compact stars were studied in the literature in the framework of $f(R)$ gravity but different equations of state were used \cite{Cappoziello4,Cappoziello3,orellana,Rescoa}. Comparing our results with those given in \cite{orellana}, we see that no agreement is found. One reason is that the polytropic exponent was not given in the reference and the parameter $\beta$ of $f(R)$ gravity was absorbed in the value of Ricci scalar $R$. The authors in \cite{Cappoziello4} pointed out that there is no self-consistent final explanation for compact objects with masses larger than OV limit. Furthermore, it is not possible to derive the mass-radius relation for compact stars from observations because measuring the radii of such objects is still a challenging task \citep{lattimer}.
\section{Conclusion}
In this work, we used one of ETG, the $f(R)$ gravity to extend the study of electrically charged polytropic spheres in the context of the Einstein-Maxwell theory. We assumed that the spheres contain a spherically symmetric distribution of charged perfect fluid. The charge density is proportional to the energy density and the charged perfect fluid is assumed to obey polytropic equation of state. Our work focused in the study of the dependence of radius to mass ratio of compact spheres with the perturbation parameter $\beta$ for some fixed values of the charge fration $\alpha$ and the polytropic exponent $\gamma$. 
For $n<5$, as in Refs. \cite{Cappoziello4, fronsdal}, the polytropic star has finite radius, its bulk is compatible with regularity at the center and that the density of the star tends to zero at some finite value of its radius. These motivated the choice of the two values $n=3$ and $n=4$ corresponding respectively to $\gamma=\frac{5}{4}$ and $\gamma=\frac{4}{3}$ in this work.

One main difficulty encountered in this work is the lack of references related to our work. One reason is that the use of different equations of state in studying compact stars that give different results. On the other side we have few information regarding cosmological data of the evolution of these stars.  We found that the radius-mass ratio ranges approximately from $0.3$ to $0.5$ in the case of $\gamma=\frac{5}{4}$. Whereas for $\gamma=\frac{4}{3}$ it ranges from $0.015$ to $7.2$.

One perspective of this work is to search for 1) extended OV limit theoretically to compare
our results and 2) the extended Buchdahl limit in $f(R)$ gravity in order to set the conditions of stability of these compact charged stars. It is also interesting to study the charged
compact stars in the freamework of ETG by considering other equations of state.
\appendix*
\section{DIMENSIONLESS RELATIVISTIC EQUATIONS OF A POLYTROPE}
For the purpose of numerical calculations, the relativistic equations of a polytrope must be written in dimensionless form. For this, we introduce the dimensionless radial coordinate given by
\begin{eqnarray}
\label{262}
&&r ={\frac {\epsilon}{\sqrt {4\pi\,\rho_{{{\it cr}}}}}},\qquad q \left( r \right) ={\frac {{\epsilon}^{2}u \left( \epsilon \right)}{\sqrt {4\pi\,\rho_{{{\it cr}}}}}},\\
\label{264} &&m  ={\frac {v \left( \epsilon \right)}{\sqrt {4 \pi\,\rho_{{{\it cr}}}}}},\qquad p \left( r \right) =\omega\,{\rho_{{{\it cr}}}}^{\gamma}\theta \left( \epsilon \right), \\ 
&& R(r)  =4\,\pi\,\rho_{{{\it cr}}}R \left( \epsilon\right), \; w(r)  =\sqrt {4 \pi\,\rho_{{{\it cr}}}}w \left(\epsilon \right).\nonumber
\end{eqnarray}
With these parametrizations, the dimensionless equations system is:
\begin{eqnarray}
\label{261}
&&{\frac {\rm d u}{{\rm d}\epsilon}} ={\frac {
\alpha\,{\theta}^{\frac{1}{\gamma}{}}}{\sqrt {a}}}-{\frac {2 u}{\epsilon}}
\\
&&v=\dfrac{\epsilon}{2} \left( {\epsilon}^{2}{u}^{2}-a+1 \right) 
\\
&&{\frac {\rm d\theta}{{\rm d}\epsilon}}\! =\!-\!\dfrac{w\theta}{2}\,
\!-\!{\frac {{\rho_{{{\it cr}}}}^{1-\gamma}\theta^{\frac{1}{\gamma}}w}{2\omega}}\!+\!{\frac {\alpha\,u{\rho_{{{\it cr}}}}^{1-\gamma}\theta^{\frac{1}{\gamma}}}{\omega\,\sqrt {a}}}
\end{eqnarray}\begin{eqnarray}
&&\frac{\rm d R}{d\epsilon}=\frac{-1}{2 \pi \epsilon \rho_{cr} \beta a \left( w \epsilon+4\right)}
\left[2 \epsilon^{2} \theta \rho_{cr}^{\gamma-1}\omega
\right.\nonumber\\&&\qquad\left.
 -a \left(4 \pi \rho_{cr} R \beta+1\right)\left(w\epsilon+1\right)+\beta \pi \rho_{_{cr}} R^{2} \epsilon^{2}
\right.\nonumber\\&&\qquad\left. 
+4\pi \rho_{cr}\beta R-\epsilon^{2} u^{2}+1\right]
\end{eqnarray}\begin{eqnarray}
\nonumber
&&{\frac {\rm d a}{{\rm d}\epsilon}}=\frac{-1}{\epsilon\, \left( 12\pi\,\rho_{{{\it cr}}}R\beta\!+\!3
 \right)  \left( w\epsilon\!+\!4 \right)}\left[
4\,{\epsilon}^{2} \left({\epsilon}w+4 \right) \theta^{\frac{1}{\gamma}}
\right.\nonumber\\&&\left.
\qquad+24\,{\epsilon}^{2}{\rho_{{{\it cr}}}}^{\gamma-1}\omega\,\theta +12\,a{\epsilon}^{2} \left( \pi\,\rho_{{{\it cr}}}R\beta +\frac{1}{4} \right) {w}^{2}\right. \nonumber\\
\nonumber
&&-6\, \left( \beta\,{R}^{2}{\epsilon}^{2}\pi\,\rho_
{{{\it cr}}}+ \left(\frac{{\epsilon}^{2}}{6}-4\,\beta\,\pi\,\rho_{{{\it 
cr}}}(a-1) \right) R \right.
\nonumber\\
&&\left.
-{\epsilon}^{2}{u}^{2}
-\!a\!+\!1\! \right) \epsilon\,w \!-\!12\,(\beta\,{R}^{2}{\epsilon}^{2}\pi\,\rho_{{{
\it cr}}}\!-\!{\epsilon}^{2}{u}^{2
}\!-\!a\!+\!1)\nonumber\\&&\qquad +\left(48\,\beta\,\pi\,\rho_{{{\it cr}}}(a-1)-4\,{\epsilon}^{2} \right) R ]
\end{eqnarray}\begin{eqnarray}
&&{\frac {\rm d w}{{\rm d}\epsilon}} =\frac{1}{24  \left( \pi\,\rho_{{
{\it cr}}}R\beta+\frac{1}{4}\right) \epsilon\,a}\left[4\,\epsilon \left( w\epsilon+4 \right) {
\theta}^{{1}/{\gamma}}\right.\nonumber\\
&&\qquad+24\,{\rho_{{{\it cr}}}}^{\gamma-1}\omega\,\theta
\,\epsilon  -6\,  \left( \beta\,{R}^{2}{\epsilon}^{2}\pi\,\rho_{{
{\it cr}}} \right.\nonumber \\ && \left.+ \left( 4\,\beta\,\pi\,(a+1)\rho_{{{\it cr}}}+{\epsilon^{2}}/{6} \right) R-{\epsilon}^{2}{u}^{2}+
a+1 \right) w\nonumber \\
&&\qquad \left.+12\,\epsilon\,\rho_{cr} \left( \beta\,\pi\,\rho_{{{\it cr}}}{R}^{2
}+{u}^{2}+{R}/{6}\right)\right]
\label{310}
\end{eqnarray}

The boundary conditions for these dimensionless equations are at the center, for $\epsilon=0,\; u(0) = 31.62,\; v(0) = 0,\; \theta(0) = 1,\; a(0) = 0,\; w(0) = 1,\; R(0) = 0$  and at the surface of the star, for some $\epsilon_{s}$ (corresponding to the radius of the star $R_{s}$), the value of the normalised pressure $\theta$ vanishes: $\theta\left(\epsilon_{s} \right)=0$. But, for numerical purposes, we chose a certain small value of
the pressure so that the numerical program stops when the pressure becomes negative or smaller
than that chosen value: $\theta\left(\epsilon_{s} \right)=10^{-10}$. Note that the initial value of $u(\epsilon)$ can take any arbitrary value due to the form of the equation $q(r)=\frac{\epsilon^{2}u(\epsilon)}{\sqrt{4\pi \rho_{cr}}}.$

\begin{acknowledgments}
The authors would like to thank N. Bouayed for reading the manuscript and giving fruitful
comments.
\end{acknowledgments}

\end{document}